\documentclass{article}
\usepackage{arxiv}
\usepackage[utf8]{inputenc} 
\usepackage[T1]{fontenc}    
\usepackage{hyperref}       
\usepackage{url}            
\usepackage{booktabs}       
\usepackage{amsfonts}       
\usepackage{nicefrac}       
\usepackage{microtype}      
\usepackage{lipsum}		
\usepackage{graphicx}
\usepackage{natbib}
\usepackage{doi}

\title{Extracting Domain-specific Concepts from Large-scale Linked Open Data}

\author{ {Satoshi Kume} \\
	Faculty of Information and Communication Engineering\\
	Osaka Electro-Communication University\\
	Neyagawa, Osaka, Japan \\
	\texttt{k-satoshi@osakac.ac.jp} \\
	\And
	{Kouji Kozaki} \\
	Faculty of Information and Communication Engineering\\
	Osaka Electro-Communication University\\
	Neyagawa, Osaka, Japan \\
	\texttt{kozaki@osakac.ac.jp} \\
}

\renewcommand{\shorttitle}{\textit{arXiv} Kume, et al.}

\begin{document}
\maketitle

\begin{abstract}
We propose a methodology for extracting concepts for a target domain from large-scale linked open data (LOD) to support the construction of domain ontologies providing field-specific knowledge and definitions. The proposed method defines search entities by linking the LOD vocabulary with technical terms related to the target domain. The search entities are then used as a starting point for obtaining upper-level concepts in the LOD, and the occurrences of common upper-level entities and the chain-of-path relationships are examined to determine the range of conceptual connections in the target domain. A technical dictionary index and natural language processing are used to evaluate whether the extracted concepts cover the domain. As an example of extracting a class hierarchy from LOD, we used Wikidata to construct a domain ontology for polymer materials and physical properties. The proposed method can be applied to general datasets with class hierarchies, and it allows ontology developers to create an initial model of the domain ontology for their own purposes.
\end{abstract}


\section{Introduction}

An ontology is an explicit description of concepts \citep{Gruber1993}, and it is a core technology for systematising and processing target knowledge. In recent years, domain ontologies that provide specialised knowledge and definitions \citep{Niles2001,Jean2007} have been constructed for various fields, such as life sciences \citep{Masuya2021,Kume2017,Jupp2016}, biomedicine \citep{Stenzhorn2007,Luciano2011}, chemistry \citep{Hastings2012,Kushida2019,Fu2015} and engineering \citep{Ahmed2006}. Domain-specific ontologies are used to construct user-interface services for information retrieval and visualisation, as well as for integrating data from different fields. However, there is still no standard methodology for constructing a domain ontology \citep{Yun2011}. Most domain ontologies are constructed manually and curated by experts \citep{Kobayashi2018}. This conventional approach has several issues: ambiguous coverage of the target vocabulary, lack of reproducibility and of the validity of the ontology construction method, and lack of interoperability with other ontologies. Addressing these issues increases the complexity of ontology design in the early stages, especially for developers unfamiliar with semantic relations. Thus, a methodology is desired for developing an initial design of a domain ontology in an automatic and objective manner.

Linked open data (LOD) \citep{cyganiak2011, Alahmari2012} is structured in accordance with the Resource Description Framework (RDF), which describes resources on the Web \citep{BernersLee2001}. RDF-based LOD possesses structural connections between concepts based on triples consisting of a subject, predicate and object \citep{Tomaszuk2020}. Large-scale LODs---such as Wikidata \citep{Vrande2014, TURKI2019} and DBpedia \citep{Auer2007}---have been developed, and various ontologies are available in RDF format \citep{Edison2016, Salvadores2013, Cote2006}. While some researchers have focused on the reuse of existing LODs, a general method for building a domain ontology from structural knowledge of LOD has not yet been proposed.

Our aim was to develop a reproducible method for extracting conceptual relations for a target domain from LOD and for constructing an initial model of the ontology while ensuring interoperability with the LOD or other ontologies. We developed a workflow for extracting graph regions, which we applied to Wikidata as a case study. We also developed an objective method for evaluating the obtained vocabulary.

\section{Data Collection}

\subsection{Requirements for Ontology Construction}

The construction of a domain ontology requires designing the class hierarchy (i.e. is-a hierarchy) and defining each class (i.e. relations other than the is-a hierarchy) \citep{Boyce2007}. In this study, we considered the construction of a class hierarchy from LOD that is closely related to the target domain. This requires addressing two issues: identifying entities/concepts that are common junction points in LOD graph and determining the search range (i.e. the number of search steps) from these entities.

We propose an algorithm for analysing superordinate concept graphs (i.e. conceptual graph structures obtained by searching upward from the entities at the search starting points) to obtain a subset of the class hierarchy that is related to the target domain. The vocabulary (i.e. the search entities) for the target domain is given, and the common upper-level entities are identified according to the number of occurrences of the lowered search entities. Then, a graph analysis is performed based on the paths of related links, and the branch points and search ranges from each branch point are determined.

\subsection{Wikidata}

For verification, we applied the proposed method to Wikidata and selected polymer material science as the target domain. Wikidata is a collaboratively edited database provided by the Wikimedia Foundation, and its data are linked to the Web and are freely available \citep{Vrande2014}. As of November 2020, Wikidata has at least 90.68 million entries, and dump data are available in RDF format. Wikidata includes information on humans, taxons, administrative divisions, structures/buildings, events, chemical compounds, movies, astronomical objects and scientific articles. Thus, Wikidata can be regarded as a collection of knowledge in various domains. 

\begin{figure}[ht]
\centering
\vspace*{-2mm}
\includegraphics[width=10.0cm]{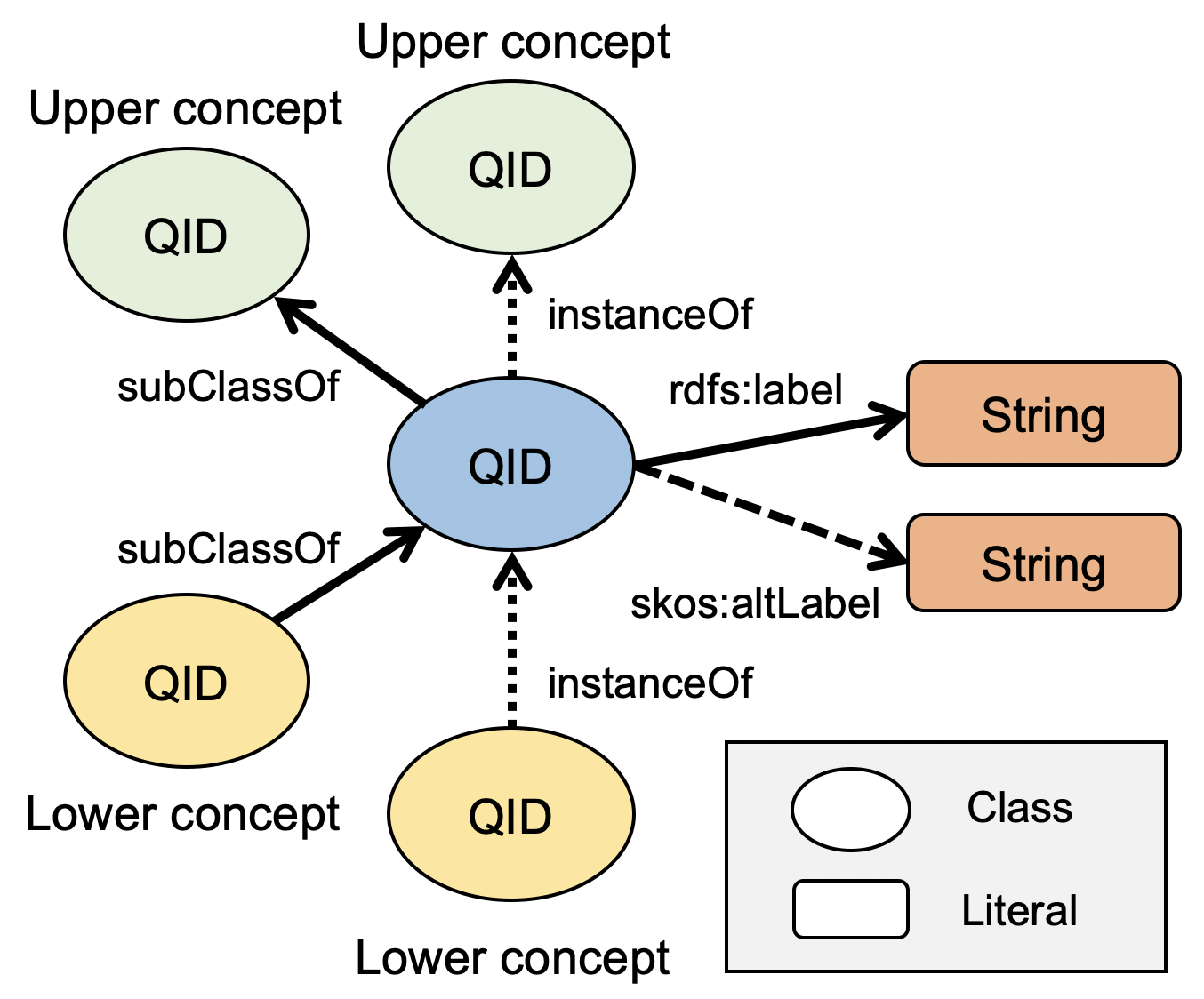}
\caption{Data model for the class hierarchy assumed in Wikidata.}
\label{fig:1}
\end{figure}

Figure \ref{fig:1} shows the data model for the class hierarchy assumed in Wikidata. Entities in Wikidata are given unique Q numbers called QIDs. The property relations of the representative label name (rdfs:label) and alias (skos:altLabel) are used to link label information. In the class hierarchy, conceptual relationships between entities are represented by the subClassOf (wdt:P279) and instanceOf (wdt:P31) properties. In Wikidata’s user-editable data model, there is no restriction preventing an entity from having or being referenced by both subClassOf and instanceOf relations. Alternatively, Wikidata allows the same entity to have both class and instance attributes, which is sometimes expressed as a loose way on the LOD cloud. For example, the entity polymer (wd:Q81163) relates with instanceOf to ’group or class of chemical substances’ (wd:Q17339814), while that same entity is also under subClassOf relations with mixture (wd:Q169336) and material (wd:Q214609). Therefore, in the present study, we assumed a data model with both property relations (Figure \ref{fig:1}). For other datasets, the flexibility of the data model is considered in the program implementation for applying the proposed method to various data models.

We conducted the experiment using a Wikidata datadump as of September 2021. There were 100,193,046 triples in the class hierarchy, and the dataset included 3,188,633 and 97,004,413 triples with the subClassOf (wdt:P279) and instanceOf (wdt:P31) properties, respectively. There were 3,813,879 triples with label information in Japanese, and the labels included 2,733,815 and 1,080,064 triples of the representative label name (rdfs:label) and alias (skos:altLabel), respectively.

\subsection{Collection of Vocabulary Related to Polymer Materials}

PoLyInfo is a polymer database in Japan that systematically organises various data related to designing polymer materials, such as polymer names, physical properties of polymers, chemical structures, measurement conditions and polymerisation methods \citep{Iimuro2007, Ishii2019}. We used the Japanese description text in PoLyInfo to obtain technical terms related to polymer materials such as polymer names, physical properties and polymerisation methods.

\section{Related works}

As preliminary research \citep{Kume2020ai}, we extracted class hierarchies using a Wikidata datadump of October 2020, which contained 94,327,098 triples in the class hierarchy. We tried to search the class hierarchies via SPARQL queries, but we found some problems. When the lower-level hierarchical structure contained more than tens of thousands of entities, searches via SPARQL queries took a considerable amount of time, and they could not completely extract all lower-level entities because of communication errors. Thus, in this study, we constructed a new workflow using text search and trimming algorithms for RDF files and then conducted experiments and evaluations using this new version of the Wikidata RDF dataset.

While previous attempts to build domain-specific ontologies and extend the concept of LOD to Wikipedia resources have been re-ported \citep{Zhang2017, Dilek2014, Khalatbari2015, Shibaki2010, Xavier2010}, this study is the first to use Wikidata and perform a graph analysis on the class hierarchy. In an additional related work, Lalithsena et al \citep{Lalithsena2017} presented an approach for extracting domain-specific hierarchical subgraphs from LOD by identifying the domain-specificity of the categories in the hierarchy. These were determined by combining different evidence using a probabilistic framework, and indicated a use case for a recommendation of movie and book domains . Aljamel et al \citep{Aljamel2015} collected distant-supervision training data from LOD for machine-learning techniques to apply information extraction for the financial domain. Unlike these methods, the proposed method provides a simple way to produce an initial model for domain-specific ontology without prior learning and categorization.

\section{Proposed Method}

\subsection{Overview of the Proposed Workflow}

\begin{figure}[ht]
\centering
\vspace*{-2mm}
\includegraphics[width=10.0cm]{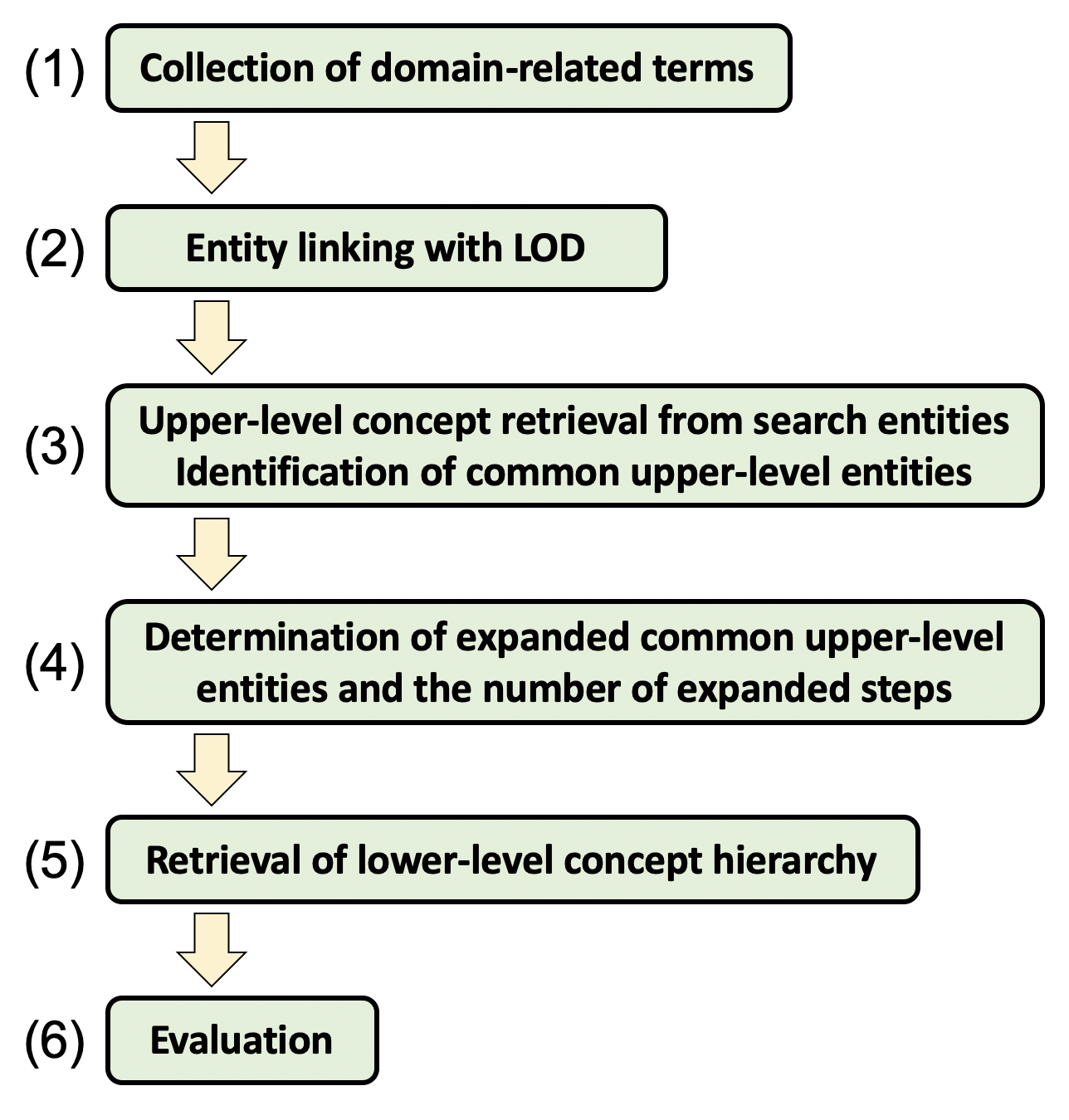}
\caption{Overview of the proposed workflow.}
\label{fig:2}
\end{figure}

Figure \ref{fig:2} shows the workflow of the proposed method. The workflow was built in an R language/RStudio environment. The programme code was implemented by using R packages such as igraph, SPARQL, purrr and plotly. An example of the workflow execution was published on GitHub as part of the agGraphSearch package (\url{https://github.com/kumeS/agGraphSearch}).

\subsection{Gathering of Domain Terms and Entity Linking with LOD}

Step 1 consists of collecting vocabulary related to the target domain in order to create a set of technical terms. The vocabulary can be extracted from domain-related documents domain or using existing keywords for the domain (e.g. domain-related web resources, indices of dictionaries and textbooks, articles, brief descriptions and keyword lists created by experts). In this study, vocabulary considered to be technical terms---such as katakana vocabulary and compound words---were picked up extensively in the manual from the Japanese description text related to the domain, which is provided as a public web resource. Further, it is possible to extract such terms using natural language processing, such as compound-word identification, but we were not too concerned about the vocabulary-gathering step. The quality of the set of search vocabulary was ensured by LOD-filtering in the next step. In practice, determining the scope of the target domain knowledge is sometimes difficult because systematic expertise is required. To address this, the proposed method selects a limited number of technical terms as described above, which are mapped to the LOD.

\begin{figure}[ht]
\centering
\vspace*{-2mm}
\includegraphics[width=10.0cm]{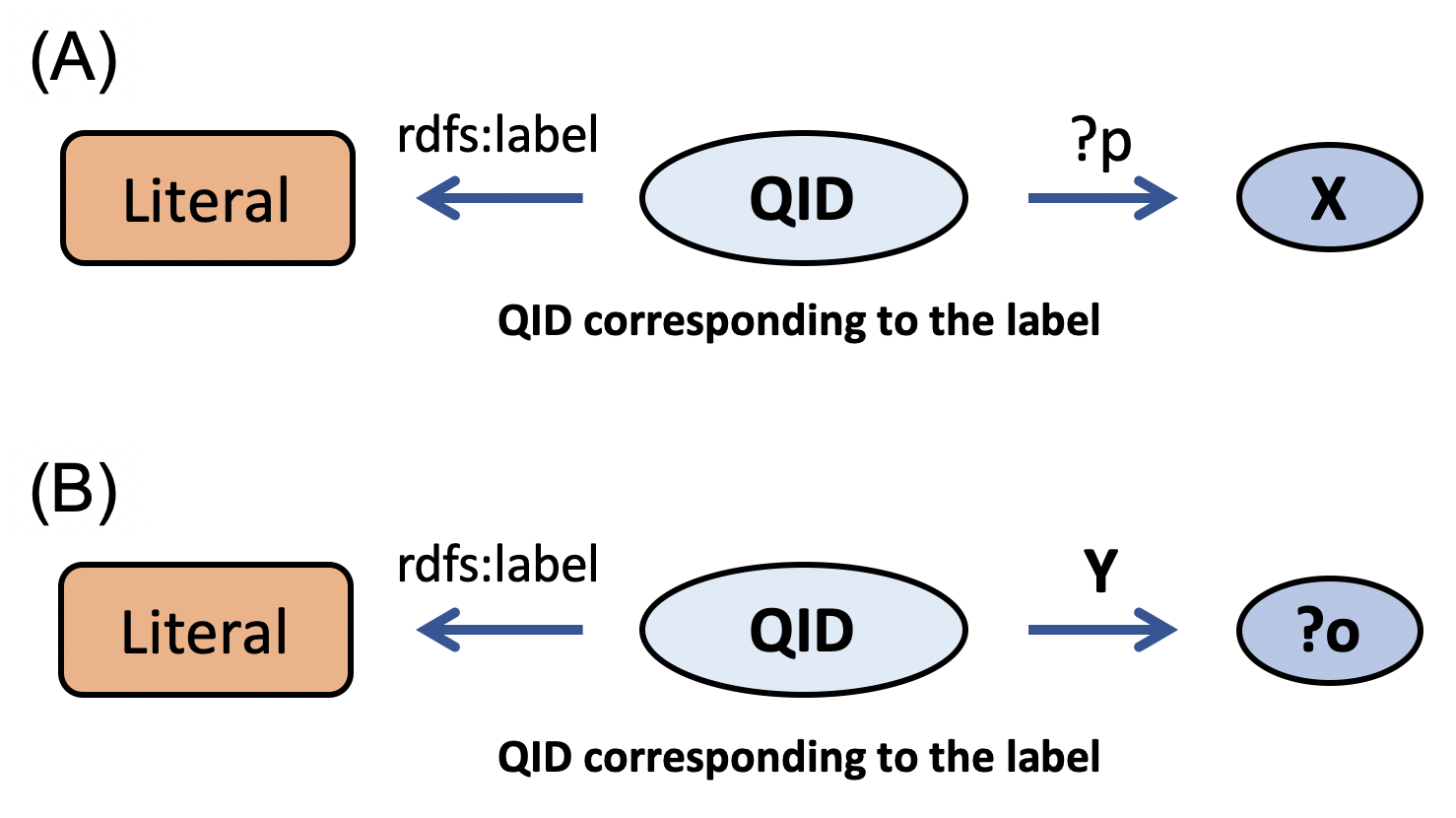}
\caption{Exclusion model for entities. (A) Exclusion of QIDs adjacent to a particular QID (X). (B) Exclusion of QIDs with a particular property (Y).}
\label{fig:3}
\end{figure}

In step 2, the collected vocabulary is linked with entities in the LOD. Although various methods of entity linking have been proposed, for our experiment we linked domain-related terms to the LOD by complete lexical matching between Japanese terms with the representative label name or alias of entities. However, this method may yield multiple entities with the same representative label name or alias. To exclude entities that are weakly related or unrelated to the target domain, we excluded QIDs adjacent to a particular QID (X), as shown in Figure \ref{fig:3}A. For example, we excluded entities related to entities such as human (wd:Q5), female name (wd:Q11879590), music term (wd:Q20202269) and movie (wd:Q11424), where ‘wd:’ is a prefix of <http://www.wikidata.org/entity/>. For example, ester (wd:Q101487) is a chemical compound consisting of a carbonyl adjacent to an ether linkage, whereas Estelle (wd:Q37080997) is a female name. However, in searching in Japanese words, these terms have the same string representation, and both entities are searched (i.e. they are homonyms in Japanese). Another example of a homonym in Japanese is silane (wd:Q410572), a chemical compound name, and Bletilla striata (wd:Q159003), a plant taxon. As previously described, Wikidata allows users to write information in a variety of domains. In several cases, this may be language-dependent. Since these representations cannot be assumed in advance, the instanceOf relation provides useful information for distinguishing them.

Another refinement is that QIDs with a particular property can be excluded, as shown in Figure \ref{fig:3}B. For example, entities with a specific property relation---such as an administrative division (wdt:P131) or gender (wdt:P21)---can be excluded, where ‘wdt:’ is a prefix of <http://www.wikidata.org/prop/direct/>. Here silane (wd:Q410572) is also a homonym in Japanese with Şiran (wd:Q390578), a district of Turkey so we used the administrative division (wdt:P131) relation to distinguish them. In addition, when entities are the subject, only entities that have a property relation of subClassOf (wdt:P279) or instanceOf (wdt:P31) are selected. The list of entities so obtained is then defined as the set of search entities.

\subsection{Retrieval of Upper-level Concepts and Identification of Common Upper-level Entities}

\begin{figure}[ht]
\centering
\vspace*{-2mm}
\includegraphics[width=10.0cm]{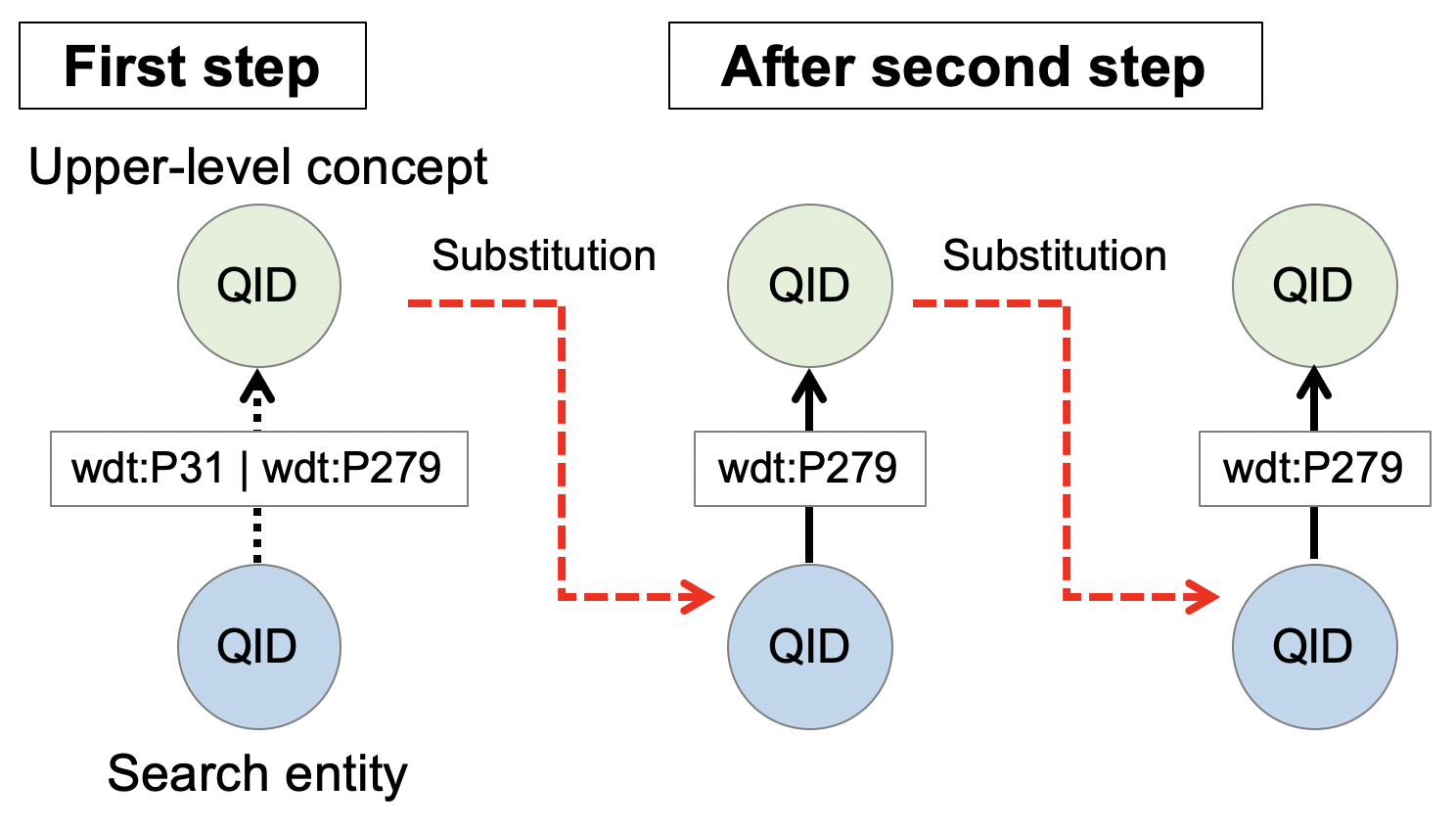}
\caption{Upper-level concept search with the data model.}
\label{fig:4}
\end{figure}

Step 3 consists in searching for relations linking each search entity to upper-level concepts based on the data model shown in Figure \ref{fig:4}. Here, the relation subClassOf (wdt:P279) or instanceOf (wdt:P31) is used to search for upper-level concepts adjacent to the search entity. For subsequent searches, only subClassOf is used to retrieve the link path recursively.

\begin{figure}[ht]
\centering
\vspace*{-2mm}
\includegraphics[width=10.0cm]{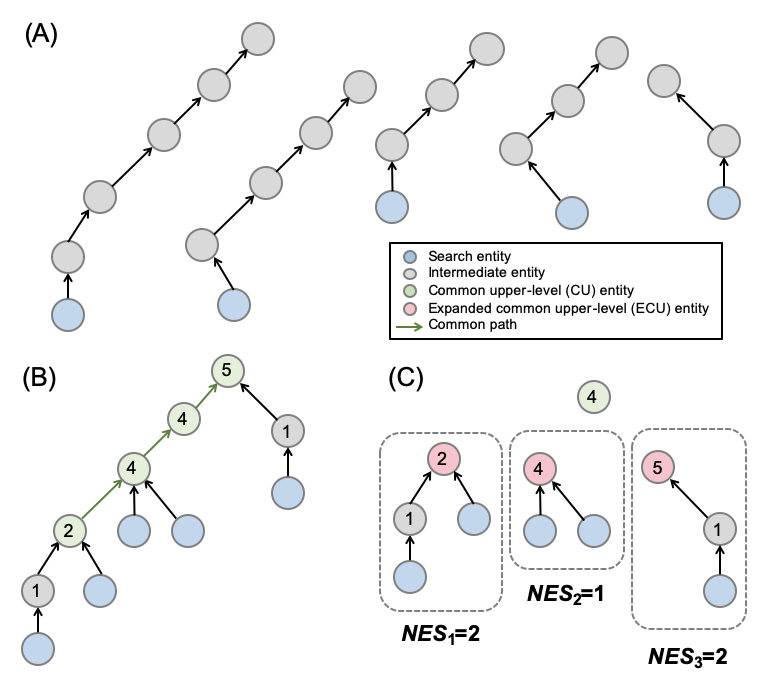}
\caption{Overview of the algorithm for upper-level concept graph integration and analysis. (A) Individual graph diagrams. (B) Integrated graph diagram. (C) Subgraph structures with each ECU entity as a vertex. The numbers in the nodes indicate the number of search entities that exist in each lower-level concept.}
\label{fig:5}
\end{figure}

This yields paths linking upper-level concepts to related search entities, as shown in Figure \ref{fig:5}A. These paths can be integrated to produce a consolidated structure of upper-level concepts with the individual graphs, as shown in Figure \ref{fig:5}B. In this graph structure, CU entities are identified by counting the number of unique paths from the search entities using the individual graphs. The number of search entities below a CU entity is consistent with the number of neighbouring links concatenated from the search entities (i.e. the values in the nodes shown in Figure \ref{fig:5}B). 

CU entities are defined as those with two or more search entities appearing in their corresponding lower-level concepts. Thus, a CU entity emerging in the graph and its superordinate concept can be a CU entity (i.e. a chain of CU entities). The path among those CU entities is called a common path. The common path is defined as the directed connections between CU entities identified in the integrated graph; this is the linkage where path duplication was observed along with the graph integration.

\subsection{Determination of the Expanded Common Upper-level Entities and the Number of Expansion Steps}

Step 4 consists in performing a graph analysis among CU entities to determine the starting points for subordinate expansion, which are ECU entities. The search range from each ECU entity (i.e. the number of steps from an ECU entity to a search entity) is then calculated. The search range of an ECU entity should not exceed neighbouring CU entities. To ensure the validity of this search rule, as a pre-processing step, common paths are removed from the upper-level concept graph, and the graph structure is then reconstructed. As shown in Figure \ref{fig:5}C, the upper-level concept graph is partitioned into subgraph structures where each ECU entity is a vertex. Other CU entities that are not analytically relevant become isolated nodes.

Here, ECU entities are defined as CU entities having two or more search entities in a lower-level concept of the partitioned complete structure. Then, the subordinate search range from each ECU entity is defined as the number of expansion steps (NES) for each ECU entity. The NES values are calculated by obtaining the shortest number of steps between the ECU entity and each search entity in the subgraph and defining the maximum value among those steps. Given an ECU entity (X), a set of shortest distances ($L_{x}$) from entity X to the N search entities in a subgraph are defined as $L_{x} = {L_1, L_2, L_3, ... L_n}$. The NES value for the expansion steps from entity X ($NES_{x}$; i.e. the maximum value among $L_{x}$) is expressed as follows:

\begin{displaymath}
NES_{x} = max( L_{x} )
\end{displaymath}

By determining the NES values, the corresponding search range of subsets from each ECU entity (Figure \ref{fig:5}C) is specified in the LOD.

\subsection{Searching for Lower-Level Concepts}

\begin{figure}[ht]
\centering
\vspace*{-2mm}
\includegraphics[width=10.0cm]{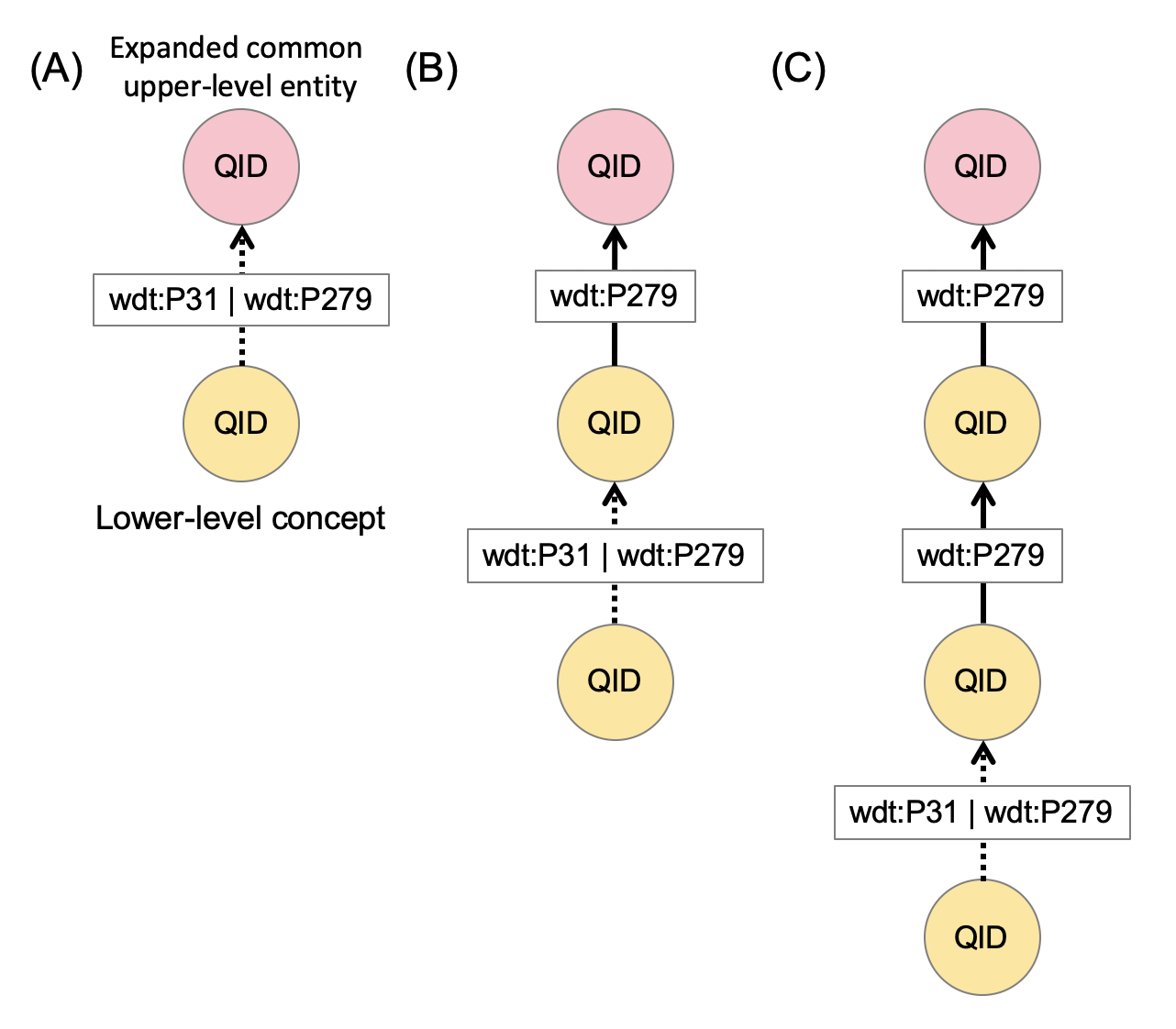}
\caption{Typical patterns of property paths for the retrieval of lower-level concepts. Models for (A) one-step retrieval, (B) two-step retrieval and (C) three-step retrieval.}
\label{fig:6}
\end{figure}

In step 5, the obtained NES is used to retrieve lower-level concepts from the ECU entities and obtain a subset of concepts in the target domain from the LOD. Concepts are retrieved from the class hierarchy by string-matching-based searching in the RDF file or SPARQL queries. Figure \ref{fig:6} shows example property paths used to search for lower-level concepts. In one-step retrieval, the subClassOf (wdt:P279) or instanceOf (wdt:P31) property is used (Figure \ref{fig:6}A). In two- or three-step retrieval, the subClassOf (wdt:P279) property is used, and the subClassOf or instanceOf property is then used at the end of the search path (Figures \ref{fig:6}B and \ref{fig:6}C). This technique produces a subset of the target class hierarchy of concepts from the LOD. Empirically, text searching is preferable for processing RDF files because network communication via a SPARQL search may not be complete because it is too time consuming or because of communication failures if the search scale is large.

\subsection{Evaluation of the Collected Vocabulary}

Step 6 involves evaluating whether the dataset extracted from the LOD is a valid candidate as a domain ontology. First, a dictionary index vocabulary must be prepared for the target domain. Correct answers are generated by judging whether these index terms are present in the LOD. By matching the obtained candidate vocabulary with the correct answers, the percentage of correct answers that can be extracted from the candidate vocabulary (i.e. the recall rate) and the precision rate can be calculated. 

\begin{figure}[ht]
\centering
\vspace*{-2mm}
\includegraphics[width=10.0cm]{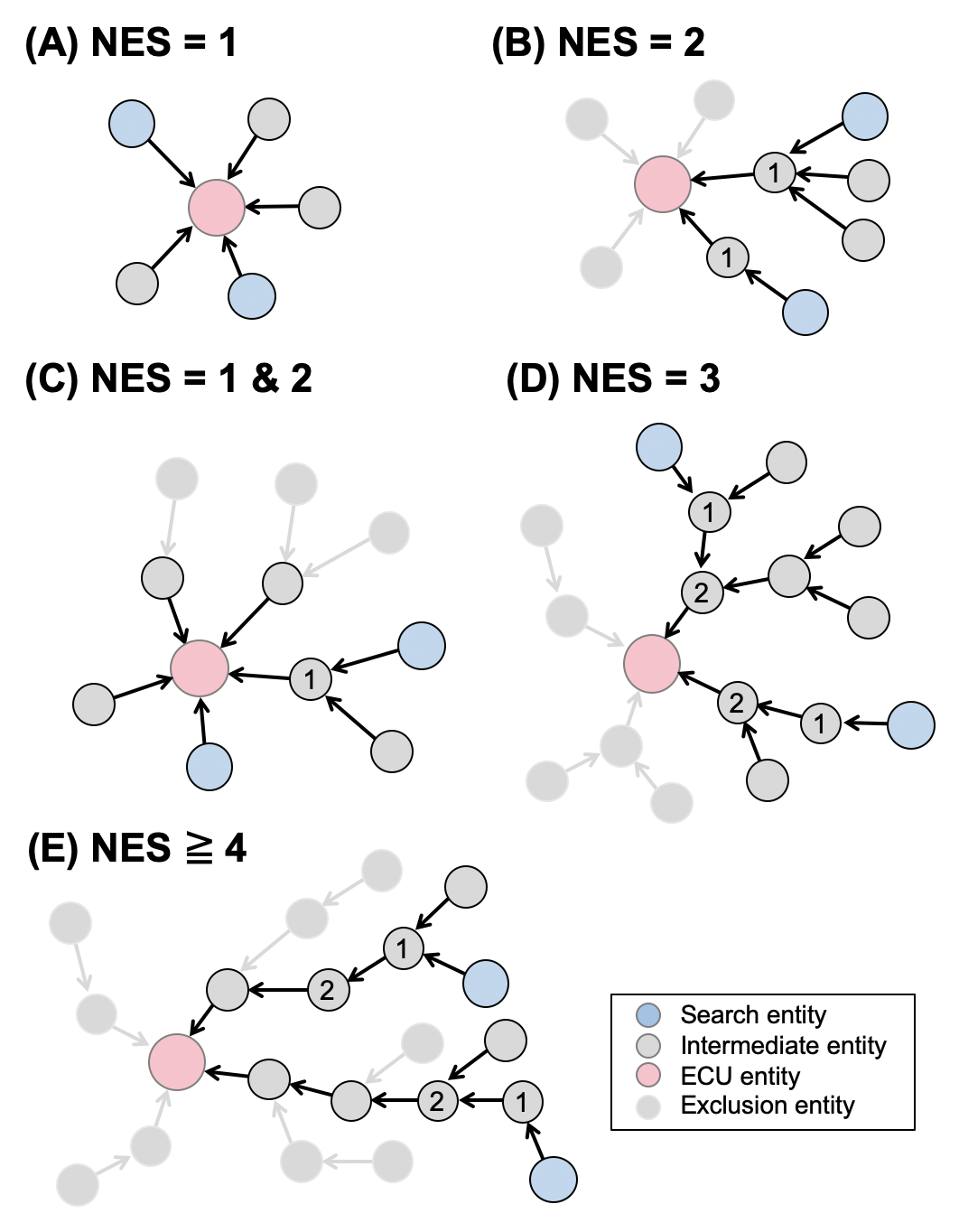}
\caption{Typical patterns of conceptual trimmed steps. Cases with (A) NES = 1, (B) NES = 2, (C) NES = 1 and 2, (D) NES = 3, and (E) NES $\geq$ 4 were represented. The number in each node shows the number of steps from each search entity. Each arrow represents a relation in the class hierarchy. The trimmed region in each graph is highlighted.}
\label{fig:7}
\end{figure}

To eliminate irrelevant entities, we examined the number of search entities in each subtree of ECU entities. Subtrees that did not contain any search entities were excluded as a post-processing step (Figure \ref{fig:7}). For NES $=$ 1, all entities were kept (Figure \ref{fig:7}A). For NES $=$ 2, only the subtrees that contains the search entity were retained (Figure \ref{fig:7}B). We also considered a case of mixed NES values (e.g. NES = 1 \& 2; Figure \ref{fig:7}C). For NES $\geq$ 3, subtrees were defined as subconcepts from two steps above the search entity in addition to the path to the ECU entity (Figure \ref{fig:7}DE). Then, the obtained candidate vocabulary was matched with the correct answers.

\section{Results}

\subsection{Case Study Using Wikidata with Polymer Materials as the Target Domain}

\begin{figure}[ht]
\centering
\vspace*{-2mm}
\includegraphics[width=10.0cm]{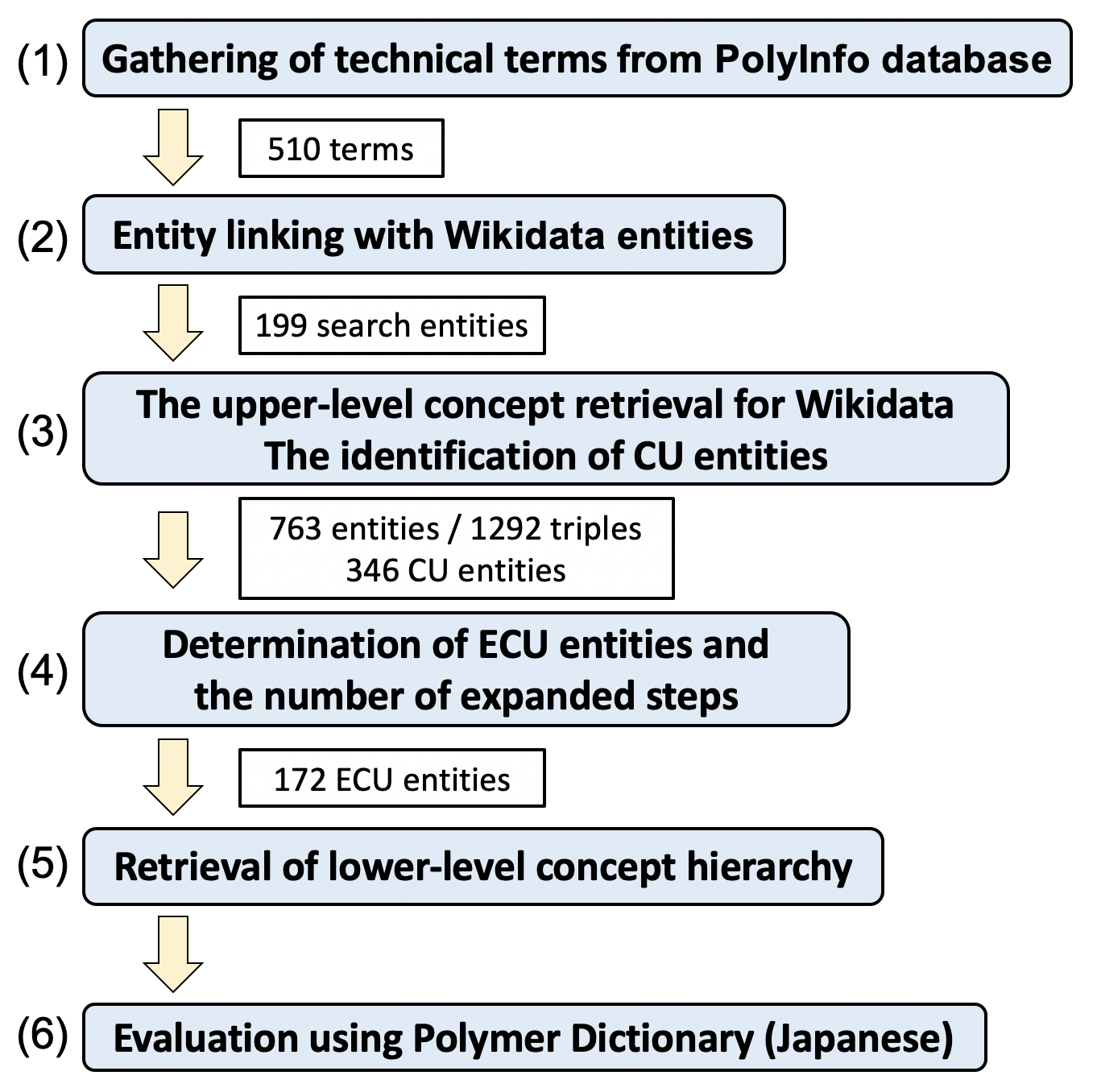}
\caption{Application of the proposed method to polymer materials.}
\label{fig:8}
\end{figure}

\begin{figure}[ht]
\centering
\vspace*{-2mm}
\includegraphics[width=10.0cm]{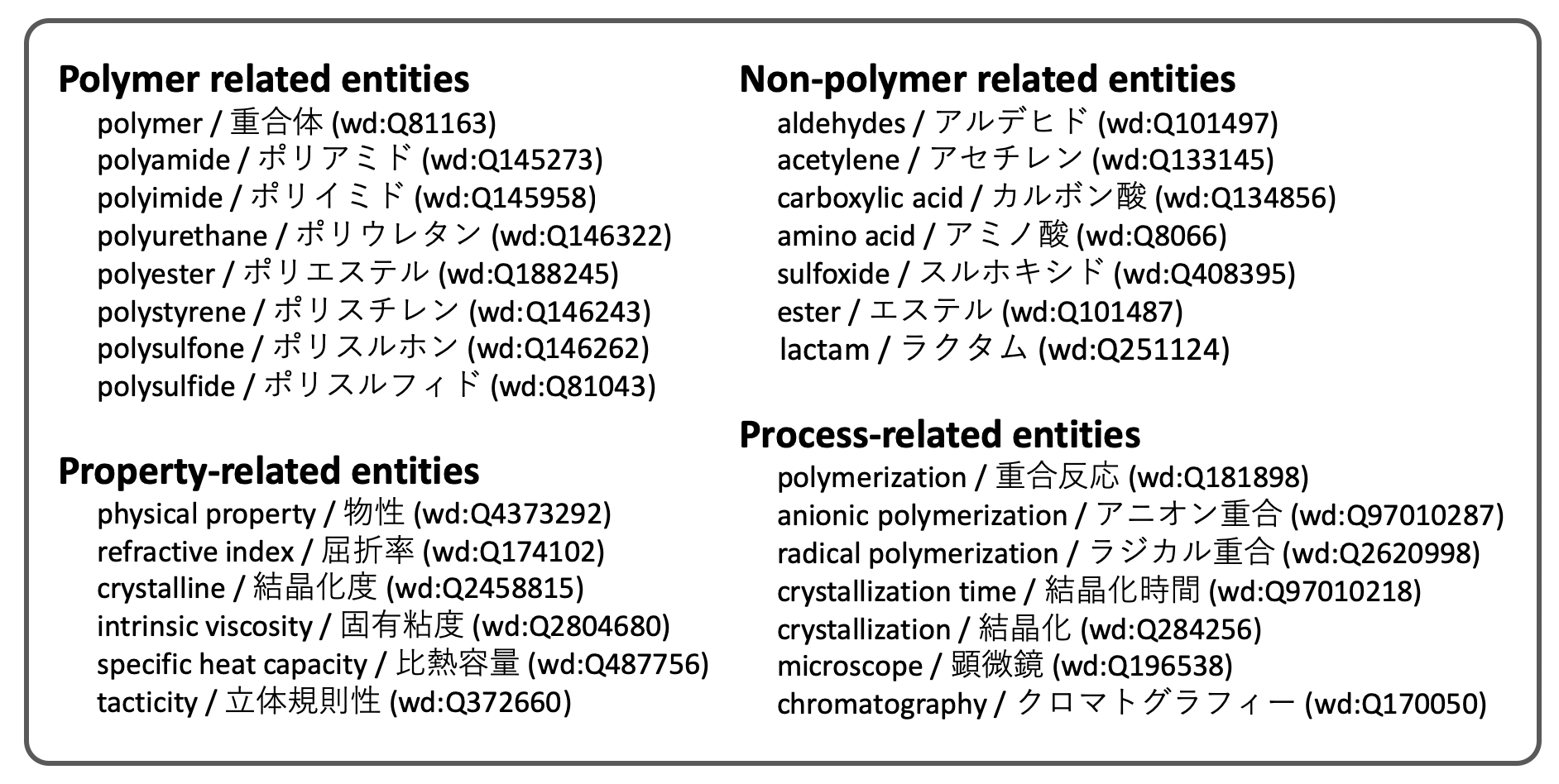}
\caption{Representative search entities obtained from technical terms for polymer materials and properties. Each item was denoted as [English term] / [Japanese term] (QIDs with the URI prefix ‘wd:’).}
\label{fig:9}
\end{figure}

\begin{figure}[ht]
\centering
\vspace*{-2mm}
\includegraphics[width=10.0cm]{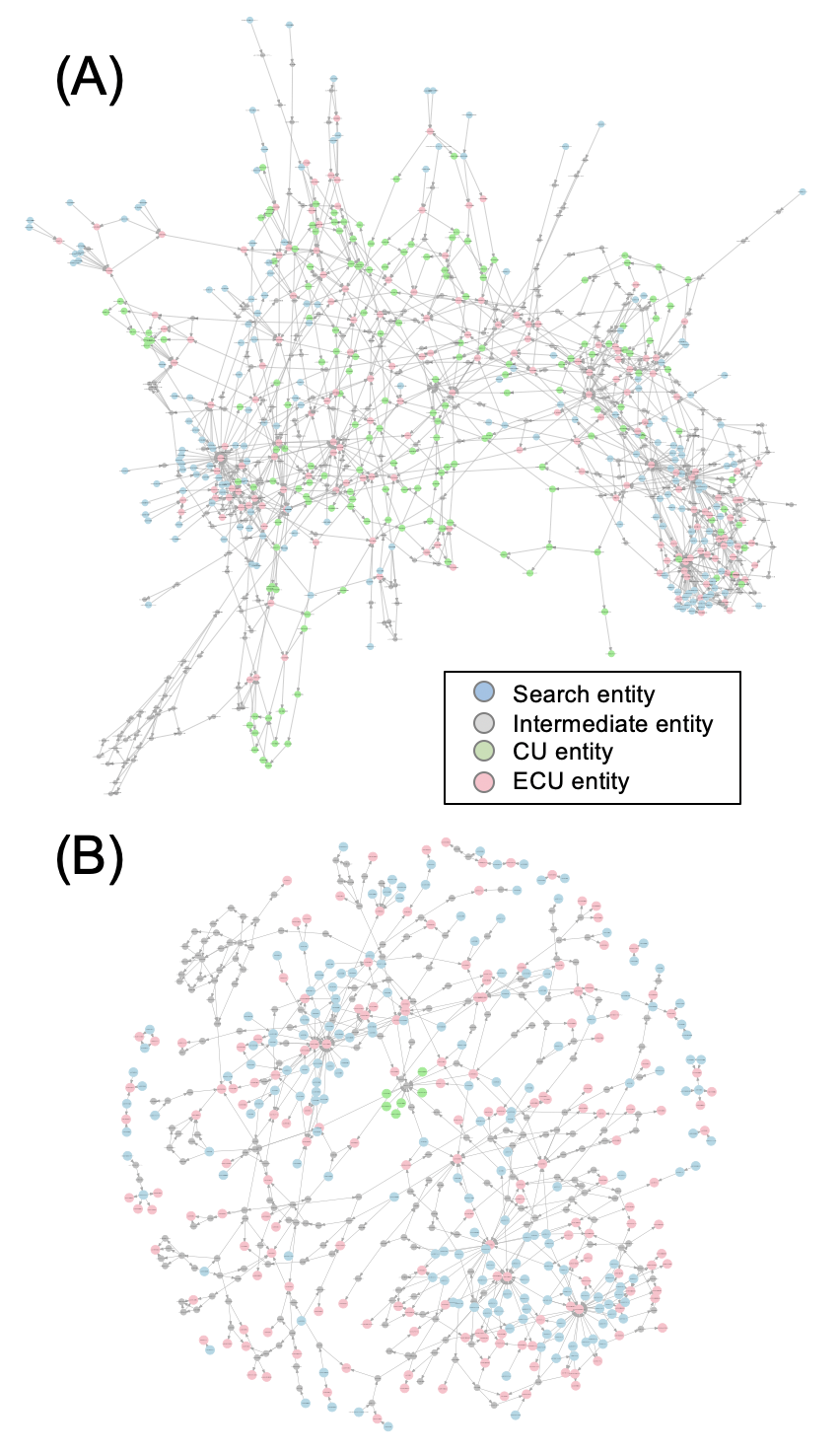}
\caption{Network diagram of an integrated superordinate graph. (A) The entire integrated superordinate graph. (B) The superordinate subgraph without the common paths between CU entities.}
\label{fig:10}
\end{figure}

\begin{figure}[ht]
\centering
\vspace*{-2mm}
\includegraphics[width=10.0cm]{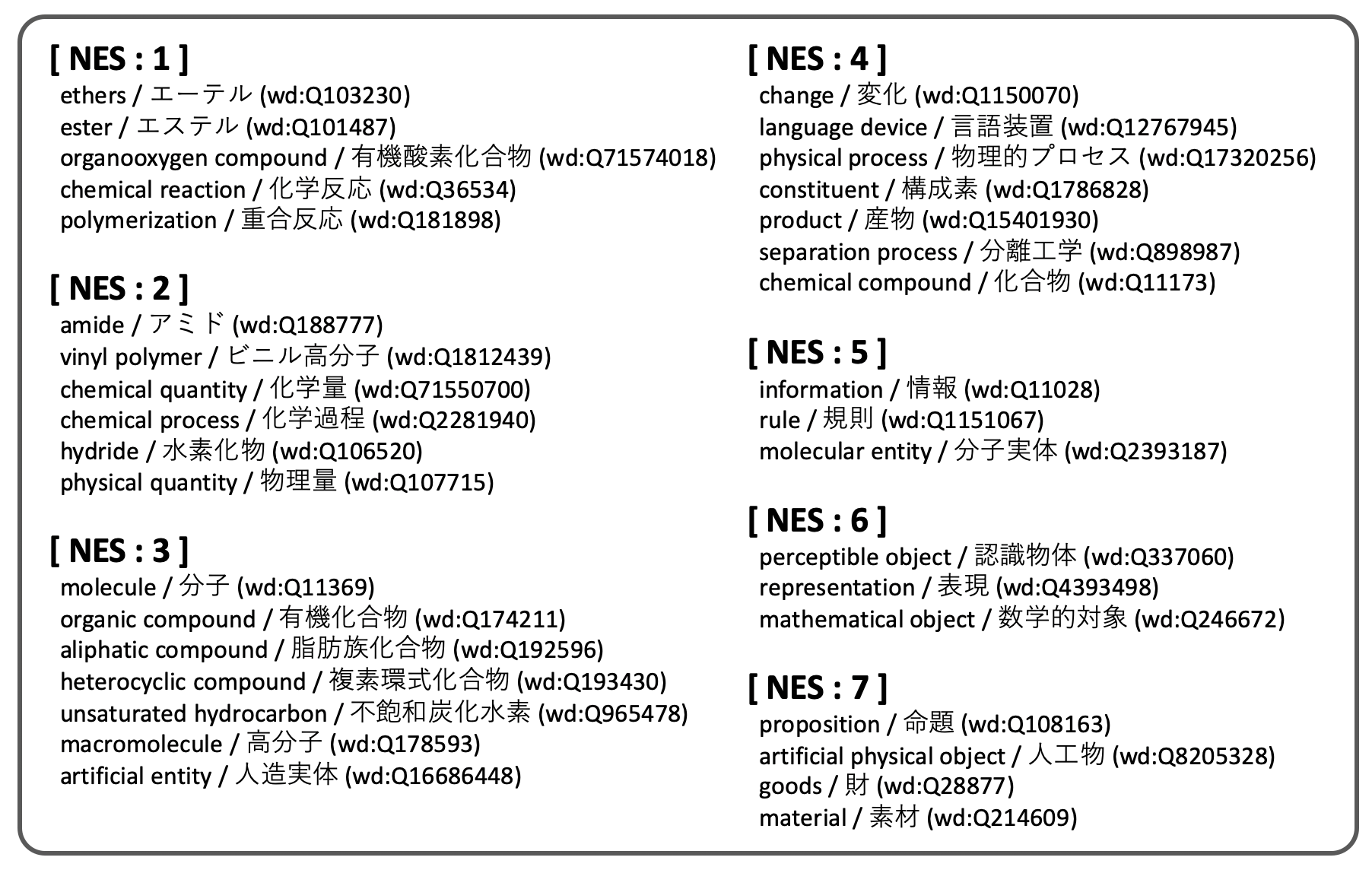}
\caption{Representative ECU entities and the number of expansion steps (NES) obtained from the technical terms for polymer materials and properties. Each item was denoted as [English term] / [Japanese term] (QID).}
\label{fig:11}
\end{figure}

We used PoLyInfo to collect lexical information related to polymer materials such as polymer names, physical properties and polymerisation methods. We manually obtained a vocabulary of 510 related terms (in Japanese) from the description text in PoLyInfo [Figure \ref{fig:8}(1)]. Entity linking between the domain vocabulary and Wikidata entities yielded 199 search entities [Figure \ref{fig:8}(2)]. Figure \ref{fig:9} shows the label names and QIDs of representative search entities, which were classified into sub-categories related to polymer materials and properties. Then, we obtained individual graphs of upper-level concepts from each search entity by exploring connections in the upward direction. We integrated the connections to obtain a graph consisting of 763 entities and 1292 triples, as shown in Figure \ref{fig:10}A. Among all the entities in the integrated graph, we calculated the path linkages between entities and identified 346 CU entities [Figure \ref{fig:8}(3)]. We then removed the common paths in the integrated graph to obtain the superordinate subgraph for further analytical purposes (Figure \ref{fig:10}B), and we used the graph-analysis algorithm described in Section 3.4 to identify the ECU entities and calculate their NES values [Figure \ref{fig:8}(4)]. This resulted in 172 ECU entities, for which Figure \ref{fig:11} shows examples of the corresponding label names, QIDs, and NES values. Technical terms related to polymers, chemical processes, and physical properties were found in ECU entities with NES = 1–4. Abstract entities, such as information and proposition were found for NES  $\geq$ 5.

\begin{figure}[ht]
\centering
\vspace*{-2mm}
\includegraphics[width=7.5cm]{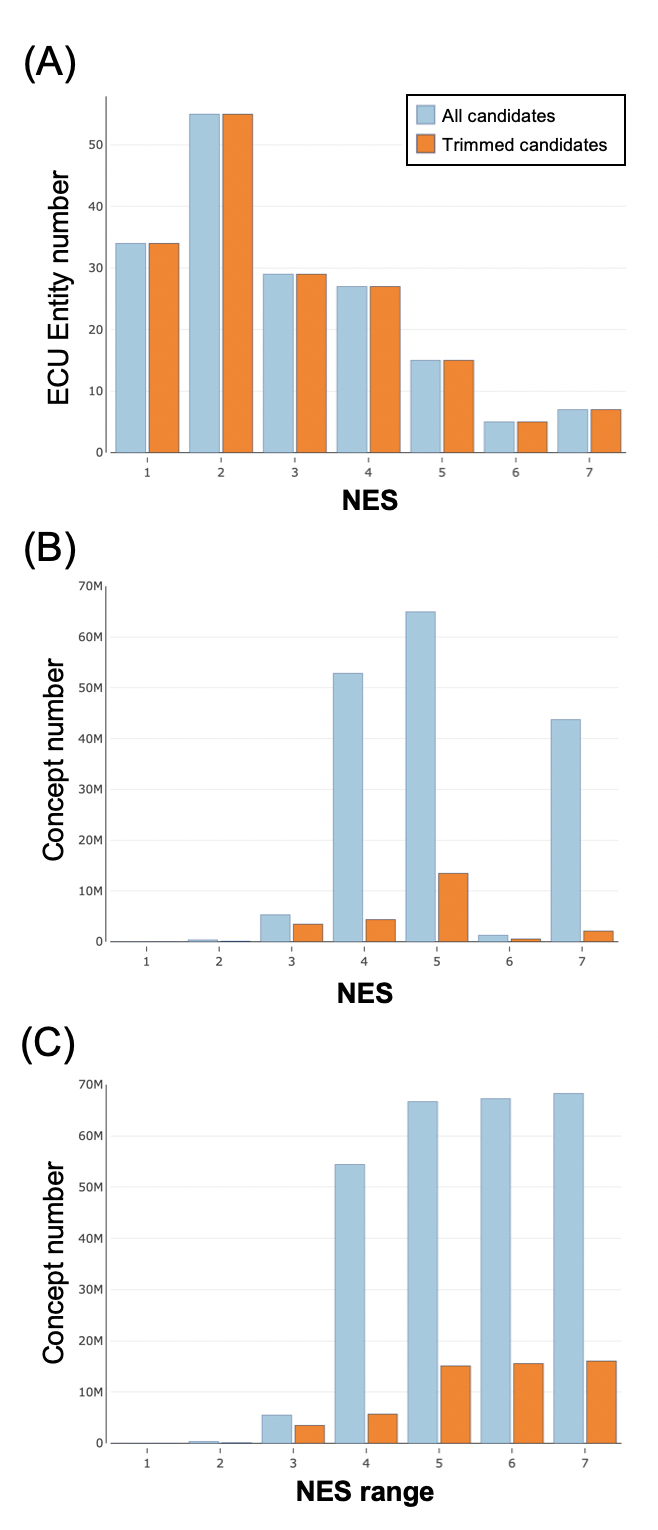}
\caption{Relationship between the number of lower-level concepts and NES. (A) Number of ECU entities at a given NES. (B) Number of concept candidates without duplicates at a given NES. (C) Cumulative number of concept candidates without duplicates with integration up to an arbitrary NES. Blue-colored bars indicate results for all candidates, and orange-colored bars indicate results for the candidates after the trimming. (B \& C) The unit of the vertical axis is 1 million.}
\label{fig:12}
\end{figure}

We then obtained lower-level concepts for each ECU entity corresponding to the obtained NES [Figure \ref{fig:8}(5)]. Among ECU entities, the largest proportion had NES = 2 (55 entities), and the number of ECU entities gradually decreased as the NES increased (Figure \ref{fig:12}A). The highest number of concepts was obtained with NES = 5, followed by those obtained with NES = 4 (Figure \ref{fig:12}B). Among the concepts obtained with NES = 4, about 1.14 million terms were concrete names for chemical compounds and were lower-level concepts belonging to the chemical-compound class (wd:Q11173). For subordinate concepts of abstract entities---such as work (wd:Q386724, NES = 5), artificial physical object (wd:Q8205328, NES = 7), creative work (wd:Q17537576, NES = 4), object (wd:Q488383, NES = 5), and spatial entity (wd:Q58416391, NES = 5)---we obtained a concept count of over 10 million. We calculated about 68.3 million unique concepts in total, of which 97.7\% had NES  $\leq$ 5, as shown in Figure \ref{fig:12}C. These 68.3 million concepts were selected as candidates for the ontology of polymer material sciences.

\subsection{Evaluation of Candidate Concepts for the Ontology}

\begin{figure}[ht]
\centering
\vspace*{-2mm}
\includegraphics[width=8.0cm]{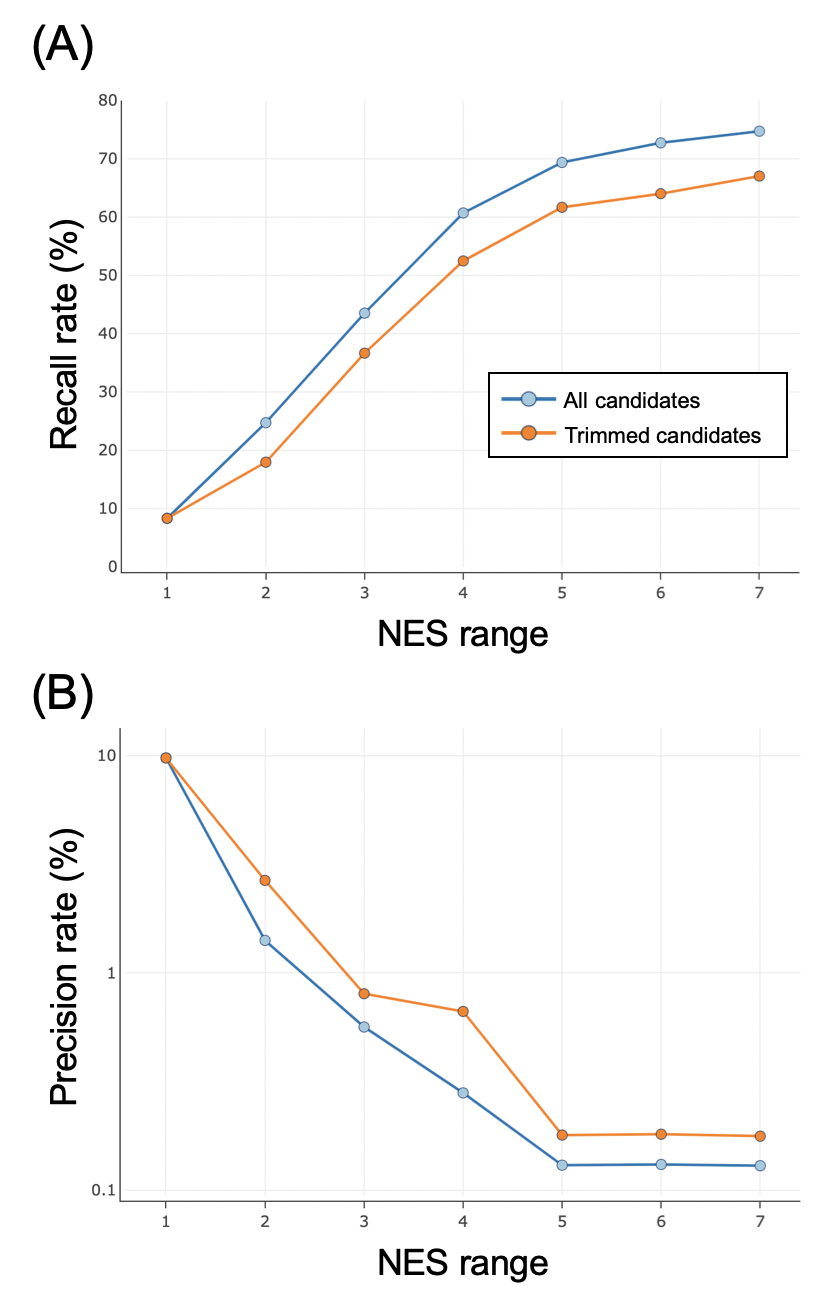}
\caption{Results for the recall rate (i.e. the proportion of correct answers extracted as candidate concepts) and the precision rate.}
\label{fig:13}
\end{figure}

The candidate concepts for the ontology were assumed to include those unrelated to the target domain. To evaluate candidate concepts for their inclusion in the ontology, we tried to optimise the NES according to the percentage of lexical occurrences in the polymer dictionary index [Figure 8(6)]. The evaluation method was based on the number of polymer-related terms (i.e. correct answers) within each NES. The polymer-related vocabulary was digitally generated by optical character recognition of Japanese terms in the Polymer Dictionary (3rd edition, edited by the Society of Polymer Science, Japan). In total, about 6700 terms were generated. Then, we referred to the vocabulary extracted from Wikidata and found 2054 matching terms, which we defined as the total number of correct answers or ground truth. We calculated the proportion of correct answers (i.e. the recall rate) and the precision rate for each concept using organised search results up to an arbitrary NES. Table 1 summarises the number of ECU entities, number of extracted candidate concepts and Japanese terms, the number of extracted correct answers, and the recall and precision rates with integration up to an arbitrary NES. The results showed that the recall rate increased with the number of candidate concepts. Figure \ref{fig:13}A plots the recall rate against the NES and shows that the recall rate increased up to NES = 5 and then plateaued at greater NES. This result is mostly consistent with the conjecture that lower-level concepts of ECU entities with a high NES contain abstract and general terms rather than technical terms. Conversely, the precision rate gradually decreased up to NES = 5. Thus, the results suggest that the candidate concepts for the ontology should contain concepts up to NES = 5.

\begin{figure}[ht]
\centering
\vspace*{-2mm}
\includegraphics[width=10.0cm]{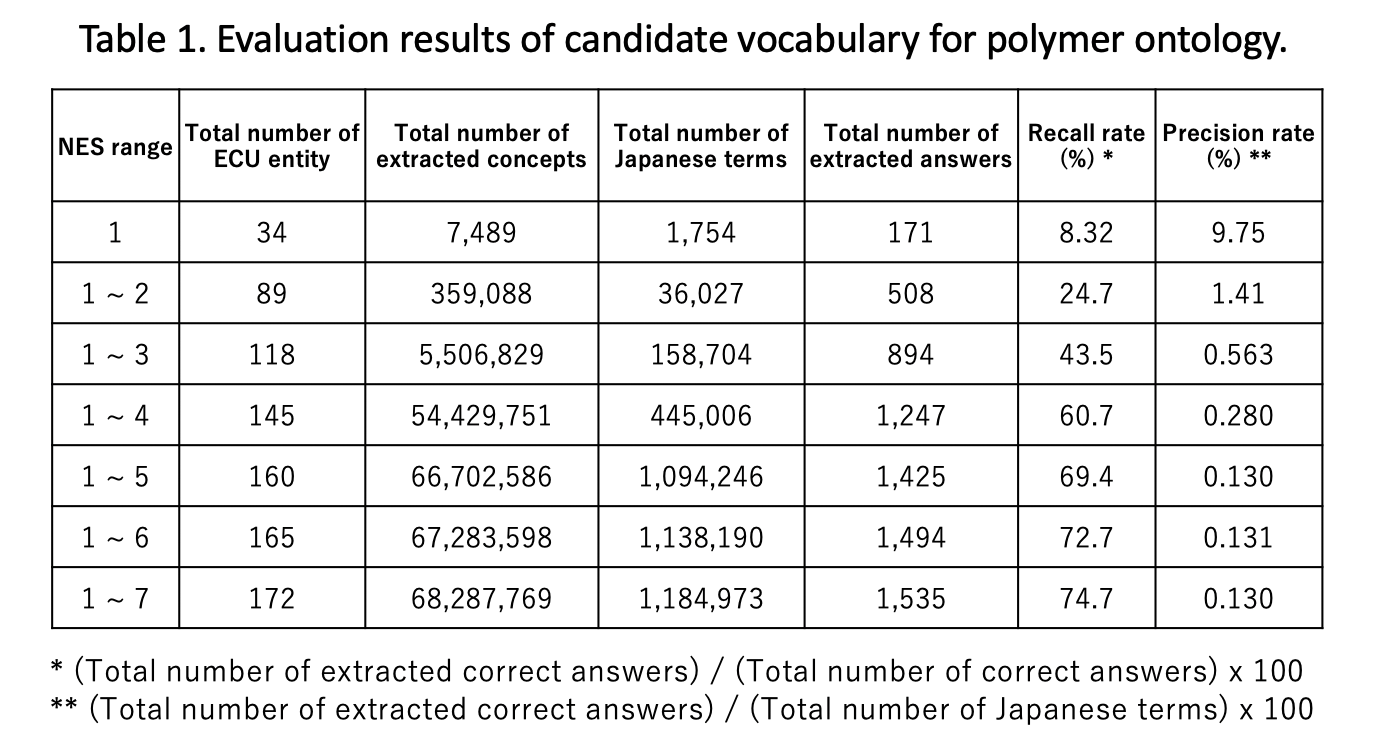}  
\label{fig:tab1}
\end{figure}

\begin{figure}[ht]
\centering
\vspace*{-2mm}
\includegraphics[width=10.0cm]{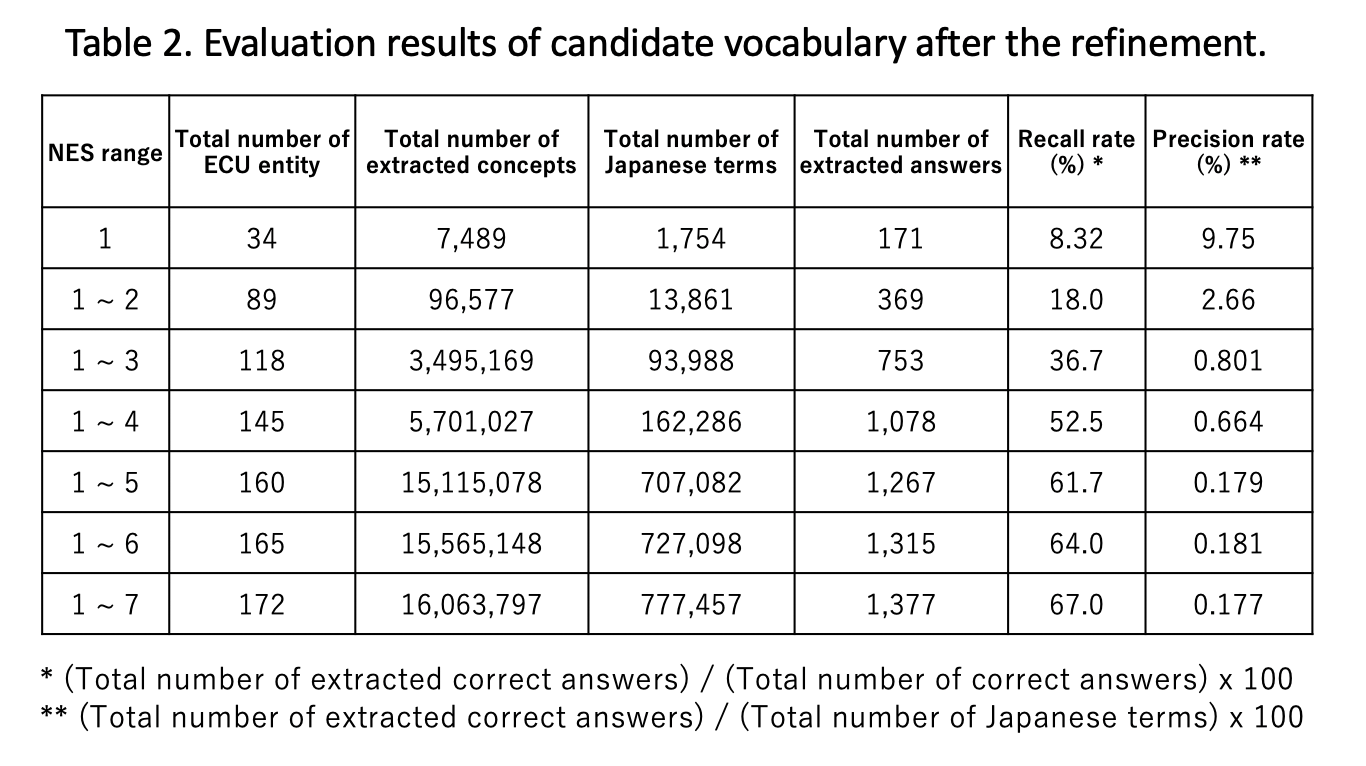}  
\label{fig:tab2}
\end{figure}

\subsection{Trimming Candidate Concepts for the Ontology Based on Search Entities}

Actual LODs such as Wikidata are created manually, and they contain unexpected class hierarchies of concepts that are not strongly related to the domain knowledge. Thus, the class hierarchy of concepts needs to be pruned as a post-processing step. Lower-level concepts are retrieved from each ECU entity and are then narrowed down based on the appearance of search entities in subtrees, as previously shown in Figure \ref{fig:7}. The number of search entities used was the same (Figure \ref{fig:12}A). After pruning, unique number of candidate concepts was reduced to about a quarter or less (Figure \ref{fig:12}BC). An evaluation was then performed using the trimmed vocabulary of candidates. Gathered up to NES = 7, the calculated recall rate was 67.0\%, and the precision rate was 0.177\% (Table 2), indicating similar trends to the case of all candidates (Figure \ref{fig:13}). The number of extracted concepts after the trimming step was 16,063,797 entities. These results suggest that pruning based on the search entities can increase the fit to the domain.

\section{Discussion}

Domain-specific vocabulary collection in various fields is not available. There is still no standardised methodology for building domain-specific ontologies. In most cases, they are constructed manually based on expert curation. In this study, we developed a methodology for constructing an initial model of a domain ontology in an automatic and objective manner, where a subset of concepts related to the target domain is extracted from LOD. In a validation experiment, we used Wikidata to construct an initial model of ontology for polymer material sciences that contained about 68 million hierarchy concepts corresponding to about 1.2 million Japanese terms.

The proposed method obtains a large number of candidate concepts for the domain ontology, so the concepts need to be pruned. We presented a trimming method based on search entities and obtained a trimmed class hierarchy that contained about 16 million concepts corresponding to about 0.78 million Japanese terms, suggesting the creation of a more domain-specific ontology. The trimming step can be improved by incorporating domain-specific named-entity recognition and contextual analysis \citep{Zhang2011, Li2020, Mysore2019}. However, the number of detected ECU entities would be increased with an increased number of input terms. Thus, the range of expansion should be limited to obtain a candidate ontology of appropriate size. In future work, we plan to survey the number of search entities and the corresponding range of lower-level expansion to determine the appropriate number of search entities. The appropriate number of search seeds needed to construct a new domain is still unknown and is a subject for future research. Because the proposed method focuses on the relationships of concepts in the class hierarchy, a methodology for extracting conceptual definitions outside the class hierarchy is also desirable.

In this study, we gathered technical terms by manual selection from the PoLyInfo database, but another option is to extract vocabulary from technical documents in the target domain \citep{Ren2012}. For the LOD mapping, semantics-matching techniques (e.g. word-embeddings-based approaches) may be useful. However, the number of candidate concepts obtained by the proposed method is considered to be sufficient for an initial trial, even with the complete lexical matching. On another note, it is important to know the limitations of the current LOD through this proposed method. An attempt to find and add concepts that do not exist in LODs such as Wikidata is also necessary in the future.

To generalise the proposed method to ontology construction, it should be validated using datasets such as the Interlinking Ontology for Biological Concepts \citep{Kushida2019}, Chemical Entities of Biological Interest (ChEBI) \citep{Degtyarenko2007} and Medical Subject Headings (MeSH) RDF \citep{Bushman2015}. In a preliminary study, we used ChEBI and MeSH RDF datasets to extract a class hierarchy related to polymer materials. The proposed method is applicable to these datasets, but it involves some conversions of the RDF data. As an alternative approach, we found that it is possible to use terms from a dictionary index as search seeds instead of manually collecting the vocabulary. Thus, the proposed method has the potential to be applied to more general datasets with class hierarchies instead of being limited to Wikidata. To generalize the methodology further is a future task requring validation using vocabularies from different domain knowledge.

Research on polymer informatics has focused on using structure-domain knowledge to reduce the cost and time of designing, generating, and manufacturing new materials \citep{Zhao2017, Zhao2018, Ashino2010, Huan2016, Cheung2008, Adams2008}. The developed domain ontology can be used to manage the large amounts of data involved in materials design and to integrate data such as the material composition, experimental process, and conditions.

\section{Acknowledgments}

This work was partially supported by the New Energy and Industrial Technology Development Organization (NEDO). The English language was reviewed by Enago (https://www.enago.jp/).

\bibliographystyle{unsrtnat}
\bibliography{references}  






\end{document}